\documentclass[10.9pt,conference]{IEEEtran}
\usepackage{amssymb,amsthm,amsmath,array}
\usepackage{graphicx}
\usepackage[caption=false,font=footnotesize]{subfig}
\usepackage{xspace}
\usepackage[sort&compress, numbers]{natbib}
\usepackage{stmaryrd}
\usepackage{xcolor}
\usepackage{mathtools}
\usepackage{float}
\usepackage{balance}
\usepackage{textcomp}
\usepackage{units}
\usepackage[most]{tcolorbox}
\usepackage{framed}

\begin{document}
\title{Towards Zero-power 3D Imaging: VLC-assisted Passive ToF Sensing}
\author{\IEEEauthorblockN{
        Faisal Ahmed\IEEEauthorrefmark{1}, 
        Miguel {Heredia Conde}\IEEEauthorrefmark{1}\IEEEauthorrefmark{2}, and 
        Paula {López Martínez}\IEEEauthorrefmark{2}
    }
    \IEEEauthorblockA{
        \IEEEauthorrefmark{1} Center for Sensor Systems (ZESS), University of Siegen, Paul-Bonatz-Straße 9-11, 57076 Siegen, Germany.\\
        \IEEEauthorrefmark{2} CiTIUS, University of Santiago de Compostela 15782, Santiago, Spain.}
        
}
\maketitle
\begin{abstract}
Passive Time-of-Flight (ToF) imaging can be enabled by optical wireless communication (OWC). The lighting infrastructure is the backbone of emerging light-based wireless communication. To this end, communication sources are used as opportunity illuminators to probe the scene, and an array of time-resolved pixels are exploited to demodulate the return, provided that the ToF camera can be externally synchronized. Our work employs a direct line-of-sight path to synchronize the camera externally. Together with the indirect path given by the reflections from the scene, this yields a bistatic configuration. Each Time-of-Flight (ToF) measurement induced a solution space of ellipsoidal shape and redefined the image formation model based on the bistatic configuration. In this demo, we showcase a passive ToF camera capable of delivering intensity and depth information in practice without emitting photons from the ToF camera. Our passive modality can eliminate built-in illumination sources, thus coping with optical power constraints, as is desired in future ToF cameras.

\end{abstract}

\section{Introduction}
Rapid advancements in the lighting industry have led to the development of visible light communication (VLC) infrastructure. Our work is motivated by the ubiquitous presence of VLC sources in indoor infrastructure. In this demo, we use the photon delays of VLC signals to develop a novel passive 3D modality, showing an enticing marriage of VLC with Time-of-Flight (ToF) imaging technology. The integration of VLC with ToF imaging opens new opportunities for a wide range of applications. 
Regrettably, this has not been capitalized on; rather, existing lighting sources have been regarded as background noise without further consideration. In this paper, we propose the use of this unrecognized sensing potential and demonstrate an operative passive ToF camera that uses VLC nodes as an opportunity illuminator. The show-and-tell demo presented in this work builds upon \cite{ahmed2022pseudo,ahmed2022demo}, which aimed to showcase the potential of the passive ToF camera.  These cameras are targeted at the rapidly growing field of indoor 3D sensing, autonomous vehicles, and industrial 3D perception and localization, for which the necessary infrastructure often exists.


\begin{figure}[ht!]
\centering
\includegraphics[width=0.46\textwidth]{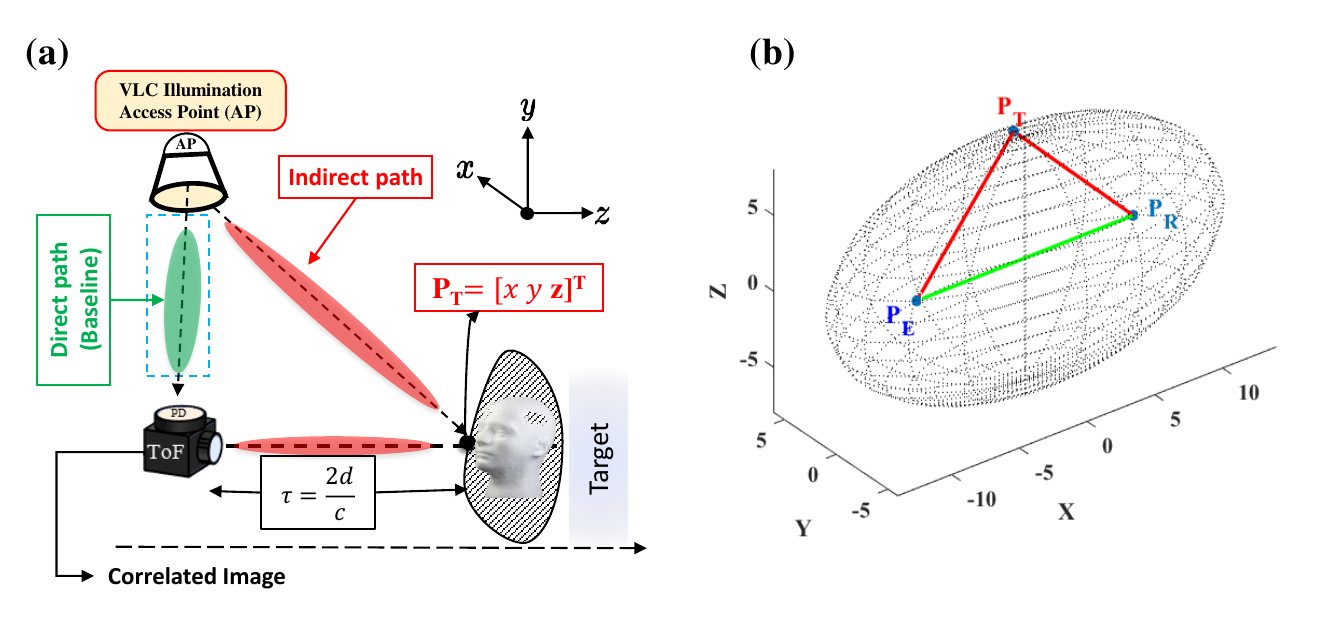} \\
\caption{(a). The geometry of the non-cooperative bistatic configuration using communication sources. (b) Resulting 3D ellipsoid model.} 
\label{fig1}
\end{figure}

A bistatic setup consists of a single emitter and two co-located receivers at different locations in 3D space, as demonstrated in Fig.~\ref{fig1}.a, enabling passive target imaging \cite{ahmed2022pseudo}. The passive pulse-based ToF approach exploits the time differences between the sequence of pulses emitted from opportunity illuminators, such as VLC sources, and their bounces captured at the ToF receiver. There are significant differences in retrieving the 3D geometry of the target with conventional ToF cameras and such a passive ToF camera. In classical ToF, the illumination source and camera are located at the same place \cite{bhandari2020one}, the target position can be determined along a single sensing path, and the camera is synchronized via an internally-generated signal. However, in a bistatic setting \cite{ahmed2022passive}, two sensing paths are exploited, yielding an ellipsoid model (see Fig. ~\ref{fig1}.b), where $ P_\mathrm{E}$, $ P_\mathrm{R}$, and $ P_\mathrm{T}$ are the emitter, receiver, and target locations, respectively. This enables us to obtain the correct target depth in a passive manner. A passive ToF camera reduces the optical power requirements, (e.g., ToF sensor IRS2381C uses \unit[1]{W} for the illumination), saving \unit[59]{\%} power consumption.


\vspace{-15pt}

\begin{figure}[ht!]
\centering
\subfloat[{Setup}]{\includegraphics[width=0.55\columnwidth]{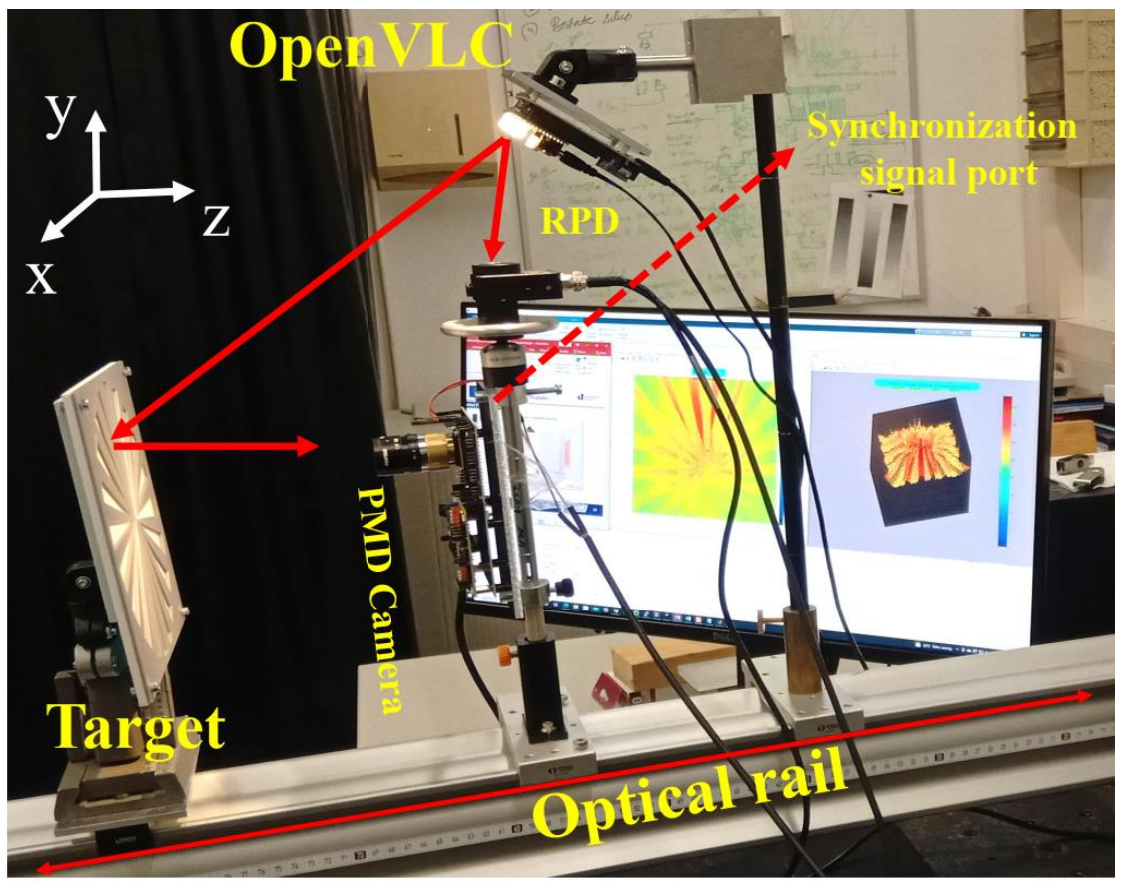}\label{fig2:1}} 
\subfloat[{Depth reconstruction}]{\includegraphics[width=0.45\columnwidth]{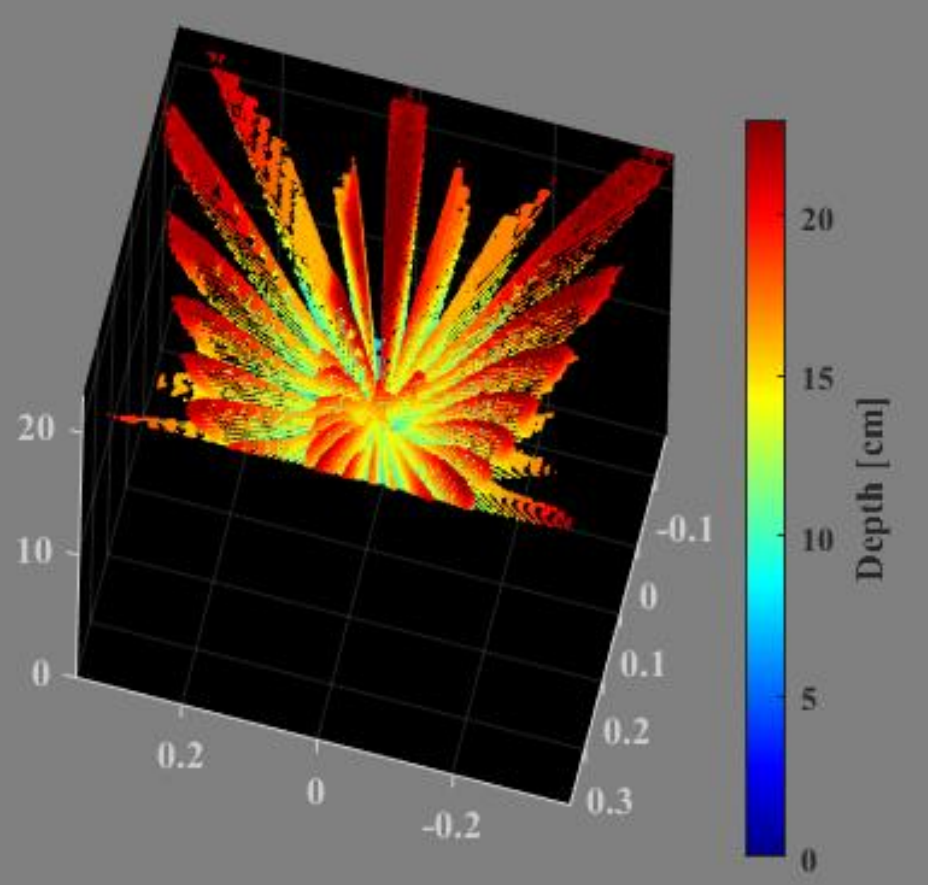}\label{fig2:2}}
\caption{(a). Experimental setup of our show-and-tell demo of a passive ToF camera. (b). Example of depth reconstruction of Boehler star from the setup.}
\label{fig2}
\end{figure}


\begin{figure}[ht!]
\centering
\includegraphics[width=0.39\textwidth]{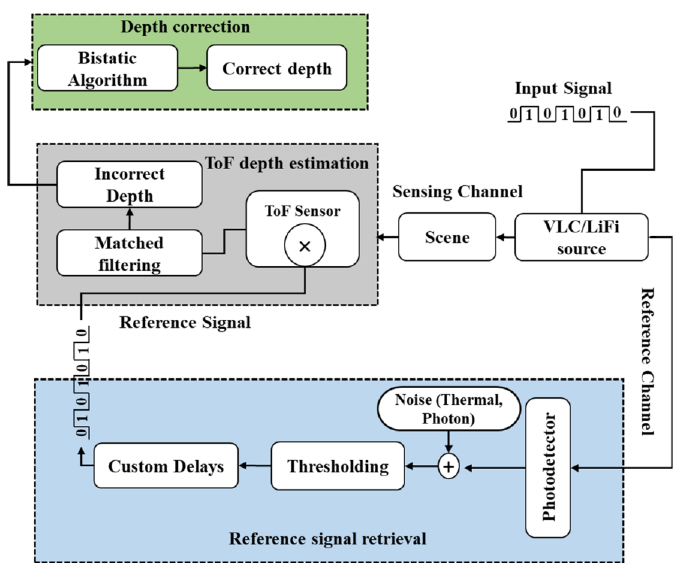} \\
\caption{Proposed computational method for passive ToF imaging.} 
\label{fig3}
\end{figure}
\vspace{-12pt}
\section{Proposed Methodology}
\textbf{System Model:} The bistatic range is determined by the flight time of the communication signal emitted from a source. The echoes are acquired at the ToF camera. In this case, the total distance between the emitter and the target can be computed. The distance between the emitter and the receiver is known as the baseline and this needs to be subtracted from the total distance. This yields a 3D ellipsoid ( Fig.~\ref{fig1}.b) in which a target satisfying the resultant bistatic range may be located. The focal points are the emitter and ToF sensor locations. Thanks to the lens calibration, we obtained the observation direction vectors for each pixel. The intersection of the ellipsoid with the direction vectors yields 3D target points and, hence, the correct depth. The 3D target location is defined as $ P_\mathrm{{T}} = P_\mathrm{{R}}+\Vec{n}_\mathrm{R} d_\mathrm{RT}$. Where $ \Vec{n}_\mathrm{R} $ is the unit vector along the observation direction and $ d_{\mathrm{RT}}$ is the correct distance between the receiver and the target.
The 3D ellipsoid is defined as, $ \mathrm{{d}} = \mathrm{d_{ET}}+\mathrm{d_{RT}}-\mathrm{d_{ER}}$. The corrected depth, $ \hat{d}_\mathrm{RT}$, is a function of the incorrect depth, $d$, and the 3D position of an emitter, $ P_\mathrm{{E}}$, as follows, 
\vspace{-5pt}
\begin{equation}
\hat{d}_\mathrm{RT} = \frac{{d_{\mathrm{ER}}}^2-d^2}{2G-2d}
\label{eq1}
\end{equation}
\noindent where $ {d_\mathrm{{ER}}}^2=\|P_\mathrm{{E}}-P_\mathrm{{R}}\|_{2}$ and $G=\langle (P_\mathrm{E} - P_\mathrm{R}),\vec{n}_\mathrm{R}\rangle$.
\textbf{Experiment:} An optical rail is used to place the setup elements as shown in Fig.~\ref{fig2}. The illumination source and the ToF camera are located at \unit[30]{cm} and \unit[20]{cm} from the target, respectively. The baseline distance between the VLC source and the photodiode is kept at \unit[10]{cm} to establish a direct link for synchronizing the ToF camera by exploiting a hardware solution labeled as \emph{reference signal retrieval} in Fig.~\ref{fig3}. The returns are projected onto the ToF pixel array using a C-mount lens with a focal length of 25 mm. The reference optical signal is launched into a thresholding circuit, after which the signal goes to a picosecond delayer to introduce custom delays for acquiring different correlation samples. The synchronization largely affects the depth accuracy. \textbf{Acquisition:} We recorded the autocorrelation function for the fundamental frequency based on emitted optical signals, which is parametrically modeled. We exploit the least squares solution or matched filtering, leveraging a fitted parametric model to obtain the time shift, and thus, the incorrect depth. The bistatic algorithm uses the incorrect depth and lens normals to accurately reconstruct the depth, accounting for the unknown source location, as we have redefined the processing pipeline shown in Fig.~\ref{fig3}. The proposed computational imaging approach is shown in Fig. ~\ref{fig3}. The final depth is calibrated to account for the mismatch of distance and the delay due to the cables. The final depth reconstruction of the Boehler star, for the setup in Fig.~\ref{fig2:1}, is shown in Fig.~\ref{fig2:2}. 

\section{Visitors Experience}
The visitors will see the 3D retrieval ability of the receiver-only ToF camera. They will observe how this passive camera is able to retrieve the 3D geometry of a challenging target. Additionally, the visitors will be briefed about how each individual component of the system works together in order to achieve such results.
\section{Conclusion}
We have introduced a pulse-based passive ToF imaging system a new computational sensing principle allowing for passive 3D depth reconstruction of the scene. We 1) developed an image formation model for such sensors, which is based on the properties of the modulated signals of opportunity illuminators, 2) proposed a hardware solution that exploits two co-located receivers, used to acquire the reference and the backscattered signals, and 3) developed a signal processing pipeline to obtain depth at the ToF sensor’s native resolution. We showcase an experimental demonstration of such capabilities using an uncontrolled illumination source.

\section*{Acknowledgment}
This project has received funding from the European Union’s Horizon 2020 research and innovation programme under the Marie Skłodowska-Curie grant agreement No: 860370 (MENELAOS\textsuperscript{NT}). This work has received funding from the Spanish Ministry of Science, Innovation and Universities under grant PID2021-128009OB-C32 and from Xunta de Galicia-Consellería de Cultura, Educación e Universidade Accreditations 2019-2022 ED431G-2019/04 and 2021-2024 ED431C2021/048 (ERDF/FEDER programme).
\bibliographystyle{IEEEtran}
\balance
\bibliography{references}
\end{document}